\documentclass[%
 reprint,
 amsmath,amssymb,
 aps,
]{revtex4-1}
\usepackage{amsmath,mathtools}
\usepackage{subfigure}
\usepackage{graphicx}
\usepackage{dcolumn}
\usepackage{bm}
\usepackage{float}

\begin{document}

\title{Kinematic constrains on interacting nucleons in Pb+Pb collisions at $\sqrt{s_{NN}}=2.76$  TeV within HIJING code}

\author{Khaled Abdel-Waged}
\altaffiliation{khelwagd@yahoo.com.}
\author{Nuha Felemban}
\affiliation{%
 Physics Department, Faculty of Applied Science, Umm Al-Qura university,\\
 P.O. Box (21955), Makkah, Saudi Arabia
}%

\begin{abstract}
The kinematic constrains on interacting nucleons in Large Hadron Collider (LHC) heavy-ion collisions are investigated in the framework of the Heavy Ion Jet Interaction Generator (HIJING) code incorporated with a collective cascade recipe.  The latter is used to implement energy-momentum conservation constrains on both primary and secondary interacting nucleons. It is found that the energy-momentum conservation constrains on the interacting nucleons affect the whole charged particle pseudo-rapidity density distribution $(\frac{dN_{ch}}{d\eta})$, at different centralities (from central (0-5\%) to peripheral (70-80\%) collisions), in Pb+Pb collisions at $\sqrt{s_{NN}} =2.76$ TeV. In particular, the kinematic constrains on the interacting nucleons are shown to reduce $(\frac{dN_{ch}}{d\eta})$ yield at mid-pseudorapidity $(\lvert\eta\rvert<2)$ in all centrality intervals, which is consistent with the LHC data. In addition, the model predicts an enhancement of the hadron production at $\lvert\eta\rvert>8$, which could be checked in future ALICE Zero Degree Calorimeter. Such an enhancement is found to be mainly due to the interactions of protons at the spectator parts of the collision. This indicates that the kinematic constrains are important for a correct geometrical treatment of  Pb+Pb collisions at LHC energies.
\end{abstract}

\maketitle


\section{\label{sec:level1}INTRODUCTION}

Microscopic hadronic cascade models, such as HIJING \cite{1,2}, EPOS\cite{3}, UrQMD\cite{4,5} and Monte Carlo Glauber-type models\cite{6,7}, have been constructed for the theoretical description of heavy-ion collisions at Large hadron collider (LHC) energies. All of them have the same basis, describing the reaction as a set of binary nucleon-nucleon (NN) collisions, which involved the primary interacting nucleons.  

At the NN-level, the primary interacting nucleons are important component for their reference role in determining the soft and hard mechanisms of initial particle production in LHC energy heavy-ion collisions. Our objective here is to address the kinematic restrictions of these interacting nucleons in the Heavy Ion Jet Interaction Generator (HIJING) framework, which is designed mainly to explore the range of possible initial conditions that may occur in LHC energy heavy-ion collisions.

HIJING \cite{1} treats NN-collision as a two component geometrical model of hard (with minijet production) and soft interactions. The hard component is characterized by a momentum transverse $(p_{T})$ larger than a cutoff scale $(p_{0})$ and is  evaluated by perturbative QCD (pQCD) using the parton distribution function (PDF) in a nucleon. While the soft interactions $(p_{T}< p_{0})$ (non-pQCD) is modelled by the formation and fragmentation of quark-gluon strings. For proton-nucleus and nucleus-nucleus collisions, HIJING implements eikonal formalism to determine the probability of collision, elastic or inelastic, and the number of jets produced in each binary collisions. 

Although the energy-momentum conservation in HIJING is
satisfied for both hard and soft mechanisms, no kinematic constrains are imposed on nucleons that are involved in the interactions. This may have consequences on the geometrical treatment of nucleus-nucleus collisions, that in turn should affect the initial particle production in heavy-ion collisions.

In this work, the  kinematic constraints on interacting nucleons are investigated by employing an improved version of HIJING (ImHIJING) with an updated  modern sets of PDFs,  and a collective cascade recipe [8]. The latter is used to implement energy-momentum conservation constrains on  both primary and secondary interacting nucleons of nucleus-nucleus collisions. We focus on the ALICE \cite{9} and ATLAS \cite{10} results of the dependence of charged particle pseudorapidity density $(\frac{dN_{ch}}{d\eta})$ on collision centrality (from 0-5\% to 70-80\%)  for Pb+Pb collisions at $\sqrt{s_{NN}} =2.76$  TeV. This is because $(\frac{dN_{ch}}{d\eta})$  provides an important information  on  initial particle production mechanisms and subsequent evolution in the created hot dense nuclear matter. Moreover,  the use of model that correctly describe $(\frac{dN_{ch}}{d\eta})$ will have important implications on the study of other phenomena such as collective flow and  jet quenching in heavy-ion collisions at LHC energies, since they depend on the initial condition of matter evolution.

The manuscript is organized as follows: in Section II, we describe the model that implements energy-momentum conservation constrains on nucleons that are involved in nucleus-nucleus interactions. Then, in Sec.III we use the model to analyze the ALICE and ATLAS results on $(\frac{dN_{ch}}{d\eta})$ as a function of collision centrality for Pb+Pb collisions at $\sqrt{s_{NN}}=2.76$ TeV. Finally, in Section IV we present our conclusions.

\section{\label{sec:level2}Description of the model}
In this section,  HIJING 1.0 code with an updated parton distribution functions (in short, improved HIJING (ImHIJING)) is supplemented with a collective cascade recipe. Aiming to establish a simple standard model, we have chosen the standard type of HIJING model.

HIJING model \cite{1,2} describes nucleus-nucleus interactions as a set of binary $NN$-collisions. At a given impact parameter ($\vec{b}$) and given center of mass energy $(\sqrt{s})$,  $NN$ scatterings are handled by the eikonal formalism. Particles produced from two colliding nucleons at high energies ($\sqrt{s_{NN}}> 4$ GeV) are described by a hard and a soft components. The hard component involves processes in which minijets are produced with transverse momentum $p_{T}$ larger than a transverse momentum cut off $p_{0}$. The inclusive cross section $\sigma_{jet}$  of the minijets is described by perturbative QCD, which depends on the parton-parton cross section $\sigma_{ab}$, parton distribution function $f_{a(b)} (x_{(a(b)}, Q^2)$ and $p_{0}$, where $x_{(a(b))}$ is the light cone fraction momentum of parton $a(b)$. The kinematics of the jets and the associated initial and final state radiation are simulated by PYTHIA model \cite{11}. On the other hand,  the soft component $(p_{T}<p_{0})$, characterized by a soft cross section $\sigma_{\text{soft}}$, treats non-perturbative processes and is modelled by the formation and fragmentation of strings,  along the lines of FRITIOF \cite{12} and DPM \cite{13,14} models.

ImHIJING  is an improved version of  HIJING1.383  in which the old Duke-Owen (DO1984) \cite{15}  parameterizations of parton distribution functions (PDFs)  are replaced by a more modern sets of Martin-Stirling-Throne-Watt  (MSTW2009) PDFs \cite{16}. Compared to DO1984 (and Gluck-Reya Vogt (GRV1995)\cite{17} parameterizations of HIJING2.0), the MSTW2009 include global fits to a larger number of  data sets, which includes both old and new types of data. The old data are improved in their precision and kinematic range. The new data include the most precise data of inclusive jet production from both HERA and Run II at the Tevatron from CDF  \cite{18} and D$\oslash$ [19, 20], that goes to larger jet $p_{T}$ values. These data are important as it constrains the gluon (and quark) distributions in the domain $0.01 \leq x \leq 0.5$ \cite{16}.

Using the MSTW2009 tabulated form of PDFs and following the same procedure as in HIJING 2.0 \cite{2}, the two free parameters of the model $p_{0}$ and $\sigma_{soft}$  are taken as energy dependent and chosen to fit $p+p (\bar{p})$ total cross sections and $(\frac{dN_{ch}}{d\eta})$ at mid-pseudorapidity.  With tuned $p_{0} (\sqrt{s})$ and $\sigma_{soft} (\sqrt{s})$, ImHIJING is found to give  the best description of $(\frac{dN_{ch}}{d\eta})$ , the multiplicity distributions of charged particles and transverse momentum spectra in non-single diffractive $p+p$ collisions at LHC energies, within the pseudorapidity interval $\arrowvert\eta\lvert<2.4$ \cite{8}.

For high energy heavy-ion collisions, both nuclear modification of nucleons \cite{8} and partons \cite{1} have to be considered. It is assumed that the parton distributions in a nucleus (with mass number A), $f_{a/A} (x_{a})$ are factorizable into parton distributions in a nucleon $f_{a/i} (x_{a})$  and the parton, $a$, shadowing factor $R_{a/A} (x_{a})$ \cite{1,2},
\begin{equation}\label{eq1}
f_{a/A} (x_{a})=A R_{a/A} (x_{a})f_{a/i} (x_{a} ).
\end{equation}

In default HIJING1.383, the shadowing effect for quarks ($q$) and gluons ($g$) is taken the same. The parton shadowing factor of the nucleon $i$($j$) from the projectile ($A$) (target($B$))  is decomposed into two parts \cite{1},

\begin{equation}\label{eq2}
R_{a/A} (x_{a},r_{i})=R_{a/A}^{0}(x_{a})-\alpha_{A} (r_{i} )  R_{a/A}^{s} (x_{a}),
\end{equation}
where
\begin{equation}\label{eq3}
\begin{split}
R_{a/A}^{0}(x_{a})&=1+1.19 \ln^{\frac{1}{6}}A[x_{a}^{3}-1.2x_{a}^{2}+0.21x_{a}]\\
&+\dfrac{1.08(A^{1/3}-1)}{\ln(A+1)}\sqrt{x_{a}} e^ {-x_{a}^{2}/0.01},
\end{split}
\end{equation}
\begin{equation}\label{eq4}
R_{a/A}^{s}(x_{a})=e^{-x_{a}^{2}/0.01}.
\end{equation}
Here $\alpha_{A} (r_{i} )$ is the shadowing parameter with $r_{i}=\sqrt{x_{i}^{2}+y_{i}^{2}} $ being the transverse distance of nucleon $i$ measured from its nucleus center.
As a result of parton shadowing, the pQCD cross section, e.g., between  two nucleons $i$ and $j$ in $A+B$ collisions becomes proportional to $\alpha_{A}(r_{i})R_{a/A}^{s}(x_{a}) f_{a/i} (x_{a}, p^{2}_{T} )\times \alpha_{A}(r_{j})R_{b/B}^{s}(x_{b}) f_{b/j} (x_{b}, p^{2}_{T}) $, which is mainly affected by the impact parameter independent shadowing parameter  $\alpha_{A} (r_{i(j)})$.

In our calculations, however, $ \alpha_{A}$ is taken as impact parameter dependence 
\begin{equation}\label{eq5}
\alpha_{A} (r_{ij} )=s_{q(g)}  (A^{1/3}-1) \: \dfrac{5}{3} \: (1-r_{ij}^{2}/R_{A}^{2} )
\end{equation}
where $R_{A}=1.2 A^{1/3}$ is the nuclear radius and  $r_{ij}=\sqrt{(b_{x}+x_{i}-x_{j} )^{2}+(b_{y}+y_{i}-y_{j} )^{2} }$  being the transverse distance of the interacting nucleon pair ($i$ and $j$),$ b_{x(y)}$ and $x_{i(j)}, y_{i(j)} $are the components of the impact parameter vector and the coordinates of the pair measured from their own nucleus. Here $s_{q(g)} $  is the shadowing parameter that should be fixed from comparison to the measured data of the centrality dependence of charged particle pseudorapidity density per participant pair of nucleons. 

The kinematic restrictions on the interacting nucleons  of nucleus-nucleus collisions are treated in ImHIJING by utilizing a collective cascade recipe \cite{21,22,8}.

At the first stage, we determine the primary interacting nucleons of the projectile ($A$) and target ($B$) nuclei by means of the eikonal formalism as implemented in HIJING.  At the second stage, we consider the non-interacting (secondary) nucleons, the spectator nucleons of the projectile/target nucleus. If the $ i^{th} $ spectator nucleon is at an impact distance $r_{ij}$ from the $ j^{th} $ primary interacting nucleon, then it is considered to be participant of the collision with the probability
\begin{equation}\label{eq6}
\varphi=C \exp(-r_{ij}^{2}/r_{c}^{2}),
\end{equation}
where $r_{c}=1.2$ fm  is the mean interaction radius and $C$ is a free parameter which determines the strength of secondary interactions. Such a nucleon can involve another spectator nucleon and so on. Note that in the case of  $C=0$, $\varphi$ reduces to the eikonal case, no secondary interactions.

The energy and momentum conservation laws are applied to the wounded (primary plus secondary interacting) nucleons through the following procedure:
\begin{enumerate}

\item
We characterize, in the case of two nuclei $A$  and $B$,  the $ i^{th} $ wounded nucleon of nucleus $A$ by the variables
\begin{equation}\label{eq7}
x_{i}^{+}=(E_{i}+p_{zi})/W_{A}^{+} \; \text{and} \; p_{Ti},
\end{equation}
and the$ j^{th}$  wounded nucleon of the nucleus $B$ by
\begin{equation}\label{eq8}
y_{j}^{-}=(E_{j}-q_{zj})/W_{B}^{-}  \; \text{and} \; q_{Tj},
\end{equation}
where
\begin{equation}\label{eq9}
W_A^+=\sum_{i=1}^{N_A}(E_{i}+p_{zi}),
\end{equation}
\begin{equation}\label{eq10}
W_{B}^{-}=\sum_{j=1}^{N_{B}}(E_{j}-q_{zj})
\end{equation}

Here $E_{i} (E_{j}) $ and $ p_{zi} (q_{zj})$ are the initial energy and longitudinal momentum of  the $i^{th} (j^{th})$ wounded nucleon. The corresponding total energy and momentum are given by$ E^{0}=\sum_{i=1}^{N_A}E_i$ and $p_{z}^{0}=\sum_{i=1}^{N_B}p_{zi}$, respectively, where $N_{A(B)}$ is the number of wounded nucleons from the projectile/target.

\item
We then ascribe to each wounded nucleon a momentum  $[\acute{x}_{i}^{+} (\acute{y}_{j}^{-} ),$ $ \acute{p}_{Ti} (\acute{q}_{Tj})]$ distributed according to the law:

\begin{equation}\label{eq11}
P(\acute{x}_i^+,\acute{p}_{Ti}  ) \varpropto \prod_{i=1}^{N_A} e^{-\frac{\acute{p}_{Ti}^2}{<p_{T}^{2}>}}
 e^{-\frac{(\acute{x}_{i}^{+}-\frac{1}{N_{A}})^2}{d^{2}}},                                                                         \end{equation}
under the constraints $\sum_{i=1}^{N_A} \acute{p}_{Ti}=0$ and $\sum_{i=1}^{N_A} \acute{x}_{i}^{+}=1$. The values of $d$ and  $<p_{T}^{2}>$ are chosen as  $0.5$ and $0.5$ (GeV/c)$^{2}$, which are fixed from the analysis of $(\frac{dN_{ch}}{d\eta})$ in $p+Pb$ collisions at  $\sqrt{s_{NN} }=5.02$ TeV \cite{8}.

\item
The final momentum of the $i^{th} (j^{th})$ wounded nucleon is obtained in terms of $(\acute{x}_{i}^{+},\acute{p}_{Ti})$  and $(\acute{y}_{j}^{-},\acute{q}_{Tj})$
\begin{equation}\label{eq12}
\acute{p}_{zi}=(\acute{W}_{A}^{+} \acute{x}_{i}^{+}-\frac{\acute{m}_{Ti}^{2}}{\acute{x}_{i}^{+} \acute{W}_{A}^{+}})/2,
\end{equation}
\begin{equation}\label{eq13}
\acute{q}_{zj}=-(\acute{W}_{B}^{-} \acute{y}_{j}^{-}-\frac{\acute{\mu}_{Tj}^{2}}{\acute{y}_{j}^{-} \acute{W}_{B}^{-}} )/2.
\end{equation}
where $\acute{m}_{Ti}^{2}={m}_{i}^{2}+\acute{p}_{Ti}^{2},\acute{\mu}_{Tj}^{2}=\mu_{j}^{2}+\acute{q}_{Tj}^{2}$, and $m_i (\mu_{j})$ is the mass of the $i^{th} (j^{th})$ wounded nucleon from A(B).

\item
We apply the energy-momentum conservation for $\acute{W}_{A}^{+}$ and $\acute{W}_{B}^{-}$:
\begin{equation}\label{14}
\begin{split}
\sum_{i=1}^{N_{A}} \acute{E}_{i} +\sum_{j=1}^{N_{B}} \acute{E}_{j}&=
\frac{\acute{W}_{A}^{+}}{2}+ \frac{1}{2\acute{W}_{A}^{+}} \sum_{i=1}^{N_{A}}\frac{\acute{m}_{Ti}^{2}}{\acute{x}_{i}^{+}}\\ &+\frac{\acute{W}_{B}^{-}}{2}
+\frac{1}{2 \acute{W}_{B}^{-}} \sum_{j=1}^{N_{B}}\frac{\acute{\mu}_{Tj}^{2}}{\acute{y}_{j}^{-}}\\
&=E_{A}^{0}+E_{B}^{0},
\end{split}
\end{equation}
\begin{equation}\label{15}
\begin{split}
\sum_{i=1}^{N_{A}}\acute{p}_{zi} +\sum_{j=1}^{N_{B}} \acute{q}_{zj}&=
\dfrac{\acute{W}_{A}^{+}}{2}-\frac{1}{2\acute{W}_{A}^{+}}\sum_{i=1}^{N_{A}}\frac{\acute{m}_{Ti}^{2}}{\acute{x}_{i}^{+}}\\
&-\frac{\acute{W}_{B}^{-}}{2}
+\frac{1}{2\acute{W}_{B}^{-}}\sum_{j=1}^{N_{B}}\frac{\acute{\mu}_{Tj}^{2}}{\acute{y}_{j}^{-}}\\
&=p_{zA}^{0}+q_{zB}^{0},
\end{split}
\end{equation}
and
\begin{equation}\label{16}
\sum_{i=1}^{N_{A}}\acute{p}_{iT}+\sum_{j=1}^{N_{B}}\acute{q}_{Tj}=0,
\end{equation}
 More explicitly, $\acute{W}_{A}^{+}$ and $\acute{W}_{B}^{-}$ are given by
 \begin{equation}\label{eq17}
\acute{W}_{A}^{+}=\frac{(W_{0}^{-} W_{0}^{+}+\alpha-\beta+\sqrt{\Delta})}{2W_{0}^{-}},
\end{equation}
\begin{equation}\label{eq18}
\acute{W}_{B}^{-}=\frac{(W_{0}^{-} W_{0}^{+}-\alpha+\beta+\sqrt{\Delta})}{2W_{0}^{+}},
\end{equation}
where
\begin{equation*}
W_{0}^{+}=(E_{A}^{0}+E_{B}^{0} )+(p_{zA}^{0}+q_{zB}^{0} ),
\end{equation*}
\begin{equation*}
W_{0}^{-}=(E_{A}^{0}+E_{B}^{0} )-(q_{zA}^{0}+q_{zB}^{0} ),
\end{equation*}
\begin{equation*}
\alpha=\sum_{i=1}^{N_{A}} \dfrac{\acute{m}_{Ti}^{2}}{\acute{x}_{i}^{+} }, 
   \beta=\sum_{j=1}^{N_{B}} \frac{\acute{\mu}_{Tj}^{2}}{\acute{y}_{j}^{-}},
\end{equation*}
and
\begin{equation*}
\begin{split}
\Delta&=(W_{0}^{-} W_{0}^{+} )^{2}+\alpha^{2}+\beta^{2}-2W_{0}^{-} W_{0}^{+} \alpha\\
&-2W_{0}^{-} W_{0}^{+} \beta-2\alpha\beta.
\end{split}
\end{equation*}

The collective cascade recipe imposes energy-momentum conservation constrains on both primary and secondary interacting nucleons in nucleus-nucleus collisions. As will be shown below, such kinematic restrictions on interacting nucleons influence the description of $(\frac{dN_{ch}}{d\eta})$, especially at mid-pseudorapidity and in the pseudorapidity region of the projectile/target.

In what follows, we denote the improvements established using the kinematic restrictions on primary and secondary interacting nucleons in ImHIJING as “ImHIJING/Primary” and “ImHIJING/Secondary”, respectively. The former corresponds to the ImHIJING calculations without secondary interactions, i.e., $C=0$, while the latter with full cascading, $C=1$, see Eq.(\ref{eq6}). In all calculations, unless otherwise mentioned, the default HIJING1.383 parameters are selected and no adjustments are attempted.
\end{enumerate}

\section{\label{sec:level3}RESULTS AND DISCUSSION}
In this section, we display the predictions of the  ImHIJING code (with and without kinematic restrictions of interacting nucleons) along with the recent measurements of ALICE (for the 30\% most central events) and ATLAS (for 40-80\% event centralities) results on $(\frac{dN_{ch}}{d\eta})$ as a function of collision centrality for Pb+Pb \cite{9,10} collisions at $\sqrt{s_{NN}}=2.76$ TeV. Because the pseudorapidity density of charged particles are measured in minimum bias (MB), we generate $7\times10^{5}$ events for a range of impact parameters from 0 to $2R_{A}$, i.e.,  7000 events are generated in equally spaced 100 impact parameter interval. We use for the different centrality classes, the range of impact parameter provided by ALICE and not the one extracted from HIJING \cite{23}. 
\begin{figure}[]
	\begin{center}
		\includegraphics[width=\linewidth]{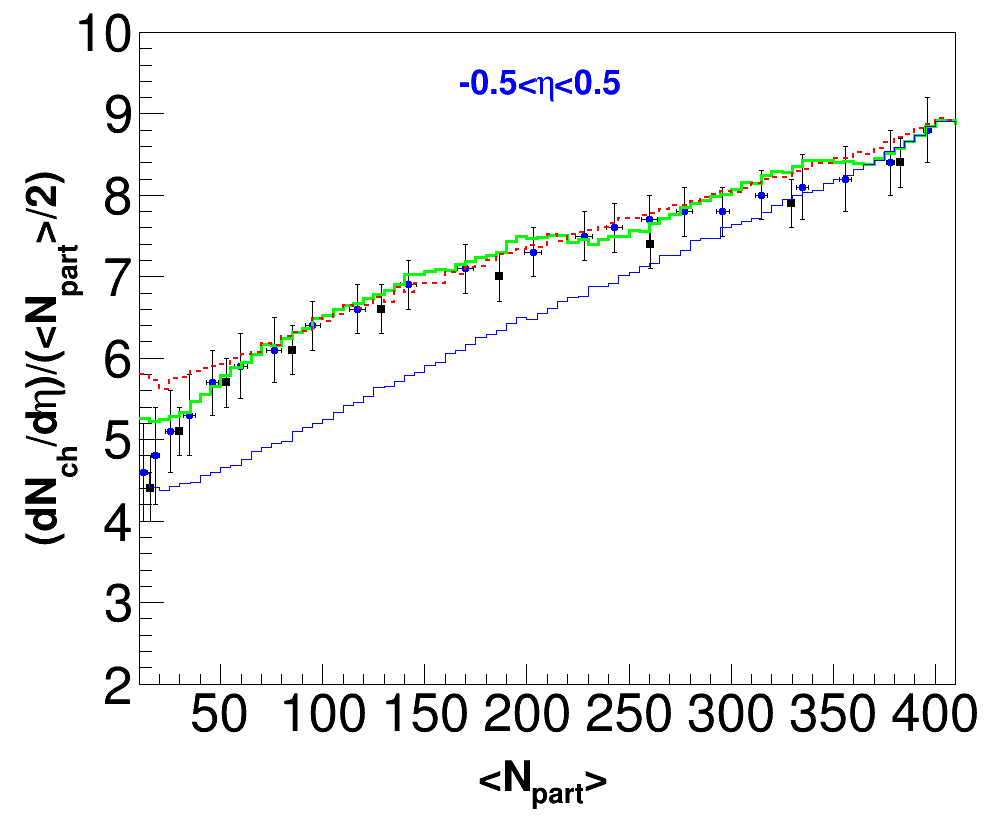}
		\caption{(COLOR ONLINE)
			Dependence of charged particle pseudorapidity density per participant nucleon pair $\frac{dN_{ch}}{d\eta}/(0.5<N_{part}>)$ on the number of participant $(<N_{part}>)$ in $Pb+Pb$ collisions at $\sqrt{s_{NN}}=2.76$ TeV. The short-dashed and thick lines denote the ImHIJING/Secondary calculation with parameter sets 1 and 2, respectively. The thin line denotes the calculation with set2 and using a fixed parton shadowing (see text). The experimental data from ALICE \cite{9} and ATLAS \cite{10} are shown by  square and closed points with error bars, respectively.}
		\label{f1}
	\end{center}
\end{figure} 
\begin{figure}[b] 
	\begin{center}
		\includegraphics[width=\linewidth]{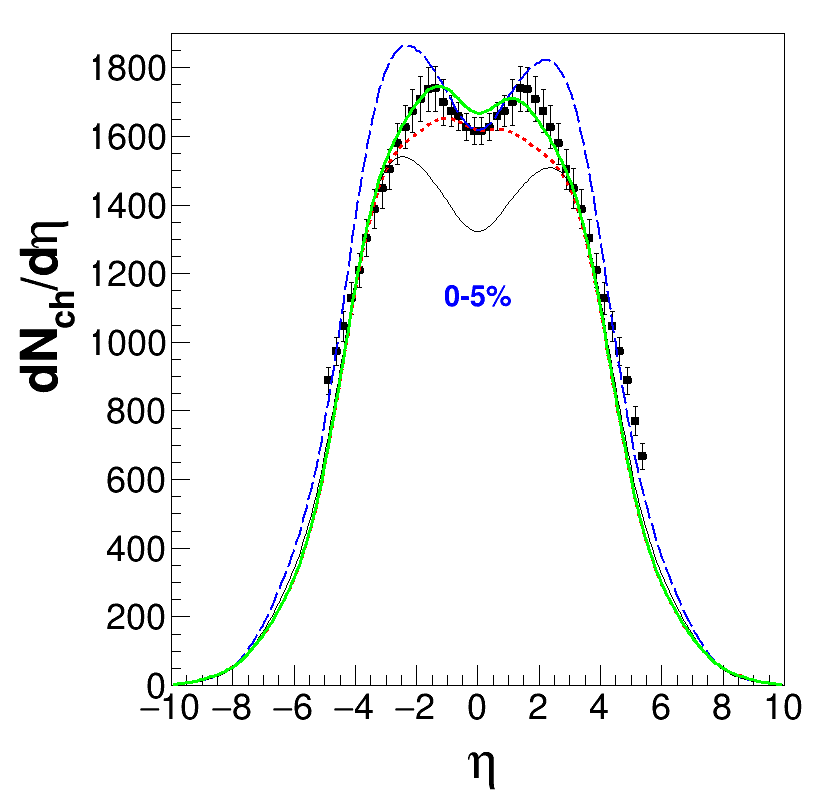}
		\caption{(COLOR ONLINE)
			Pseudorapidity density of charged particles in Pb+Pb collisions at $\sqrt{s_{NN}}=2.76$ TeV for 0-5\% centrality events. The long-dashed line denotes the ImHIJING/Secondary calculation with set 1. While the thin, short-dashed and solid lines denote the ImHIJING/Secondary calculation with set 2 and using final state interactions corresponding to jet energy loss of 0, 2 and 4.5 GeV/fm, respectively. The experimental data (square points with error bars) are from ALICE \cite{9}.
		}
		\label{f2}
	\end{center}
\end{figure}	

In order to study the effects of kinematic constrains on interacting nucleons for the reaction under study in the framework of HIJING model, one has to constrain the parton shadowing parameter $s_{q(g)}$ . In the present study, two sets of parameters are employed to demonstrate that the fitness of $(\frac{dN_{ch}}{d\eta})$  for different centrality at mid-pseudorapidity does not depend on the assumed form of $s_{q(g)}$.

The first set of parameters are fixed by keeping the value of $s_{q(g)}$ at 0.1 and  adjusting the parameters of the soft component, e.g., varying the values $a$ and $b$ of the Lund string fragmentation function $f(z) \varpropto  z^{-1} (1-z)^{a}\exp(-bm_{\perp}^{2}/z)$ here $z$ is the light cone momentum fraction of the produced hadron of transverse mass $m_{\perp}$. In particular, instead of the default values $a=0.5$ and $=0.9$ GeV$^{-2}$, that corresponds to a smaller string tension, the values $a=1.0$ and $b=0.5$ GeV$^{-2}$ are used.  These values corresponds to a larger string tension and gives a larger multiplicity density at mid-pseudorapidity than that from the default HIJING values, and thereby affect the magnitude of $s_{q(g)}$. Note that, the ImHIJING/Secondary calculations with set 1 are performed without jet quenching.         
As shown in Fig.\ref{f1}, the model results (short-dashed line) with set 1 are found to be consistent with the measured charged particle pseudo-rapidity density per participating nucleon pair $\frac{dN_{ch}}{d\eta}/(0.5<N_{part}>)$ as a function of the number of paticipants ($N_{part}$) for $Pb+Pb$ collisions at $\sqrt{s_{NN}}=2.76$ TeV \cite{9,10}.

The second set of parameters are adjusted by keeping the default soft component parameters (that is using the default Lund string fragmentation  parameters of $a=0.5$ and $=0.9$ GeV$^{-2}$) turning jet quenching on, and, finally, assuming $s_{q(g)}$ to be dependent on impact parameter. The centrality dependence of the quark/gluon shadowing, which fits the measured centrality dependence of $\frac{dN_{ch}}{d\eta}/(0.5<N_{part}>)$, takes the form 
		\begin{equation}\label{eq19}
			s_{q(g)}(c)=0.1+0.12c^{3}-0.14c^{2}-0.0003c  
		\end{equation}           
where the centrality $c$ is related to the impact parameter by the empirical formulae $c=\frac{\pi b^{2}}{\sigma_{in}}$  \cite{24} with the inelastic $Pb+Pb$ cross section $\sigma_{in}\approx 784$ fm$^{2}$ calculated from the Glauber model. Using the experimental data on deep inelastic scattering (DIS) off nuclear targets as a constraint, the  maximum value of $s_{q(g)}(c)$ (that corresponds to the most central collisions) is fixed at 0.1. According to Eq.\ref{eq19} the quark/gluon shadowing is decreasing rapidly as the impact parameter increases, with $s_{q(g)}(c)\approx0.1$ and 0.072 for $b<3.5$ fm (0-5\% centrality) and $b\leq2 R_{A}$ fm (70-80\% centrality), respectively.  As one can see, in Fig.\ref{f1}, the data could not be described by assuming only a constant $s_{q(g)}$ (thin line). The introduction of $s_{q(g)}(c)$ (thick solid line) has an effect on  the predicted $\frac{dN_{ch}}{d\eta}/(0.5<N_{part}>)$ at peripheral collisions ($<N_{part}>$ $<280$) and results in a better agreement with data. 
 \begin{figure}[] 
 	\begin{center}
 		\includegraphics[width=\linewidth]{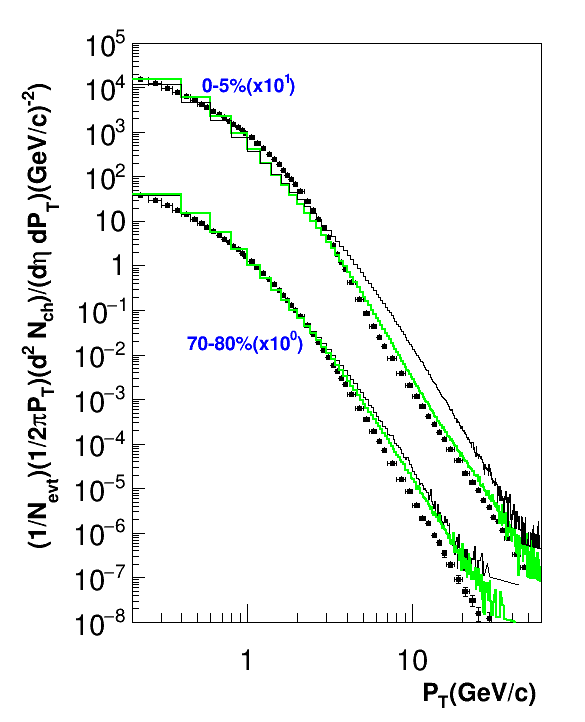}
 		\caption{(COLOR ONLINE)
 			Invariant transverse momentum distributions of charged particles at central (0-5\%) and peripheral (70-80\%) collisions in in $Pb+Pb$ collisions at $\sqrt{s_{NN}}=2.76$ TeV, from ALICE experiment [23] (points with error bars) as compared to ImHIJING/Secondary calculations with (thick lines) and without (thin lines) jet quenching. For clarity, the histograms and the data have been multiplied by the indicated values. 
 		}
 		\label{f3}
 	\end{center}
 \end{figure}   
 \begin{figure*}[] 
 	\begin{center}
 		\includegraphics[width=\linewidth]{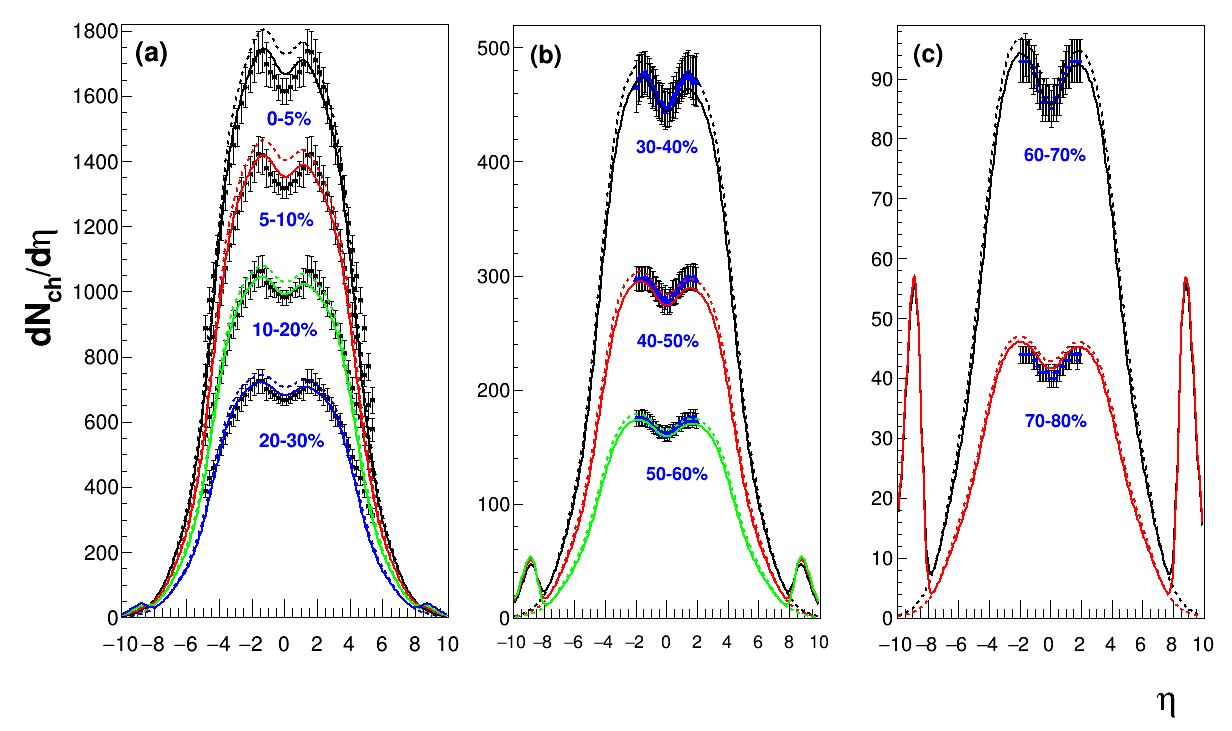}
 		\caption{(COLOR ONLINE)
 			Pseudorapidity density of charged particles in Pb+Pb collisions at $\sqrt{s_{NN}}=2.76$ TeV for nine centrality intervals. The solid and short-dashed lines denote the ImHIJING/Secondary and ImHIJING/Primary calculations, respectively. The square and solid points with error bars denote ALICE \cite{9} and ATLAS \cite{10} data, respectively. 
 		}
 		\label{f4}
 	\end{center}
 \end{figure*}  
 \begin{figure*}[] 
 	\begin{center}
 		\includegraphics[width=\linewidth]{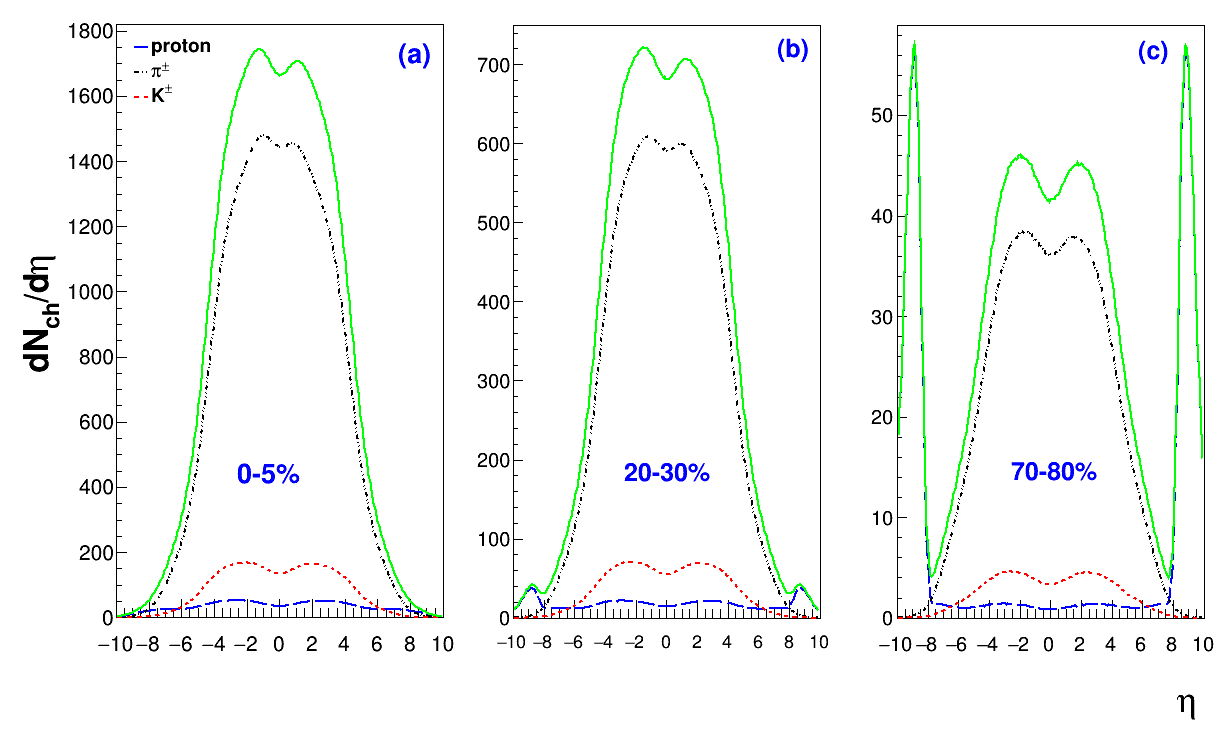}
 		\caption{(COLOR ONLINE)
 			ImHIJING/Secondary predictions of specific hadron species in $Pb+Pb$ collisions at $\sqrt{s_{NN}}=2.76$ TeV: (a) for central (0-5\%), (b) for semi-peripheral (20-30\%) and, (c) for peripheral (70-80\%) interactions. The thick lines denote predictions for total charged particles as a reference.
 		}
 		\label{f5}
 	\end{center}
 \end{figure*}
 \begin{figure*}[] 
 	\begin{center}
 		\includegraphics[width=\linewidth]{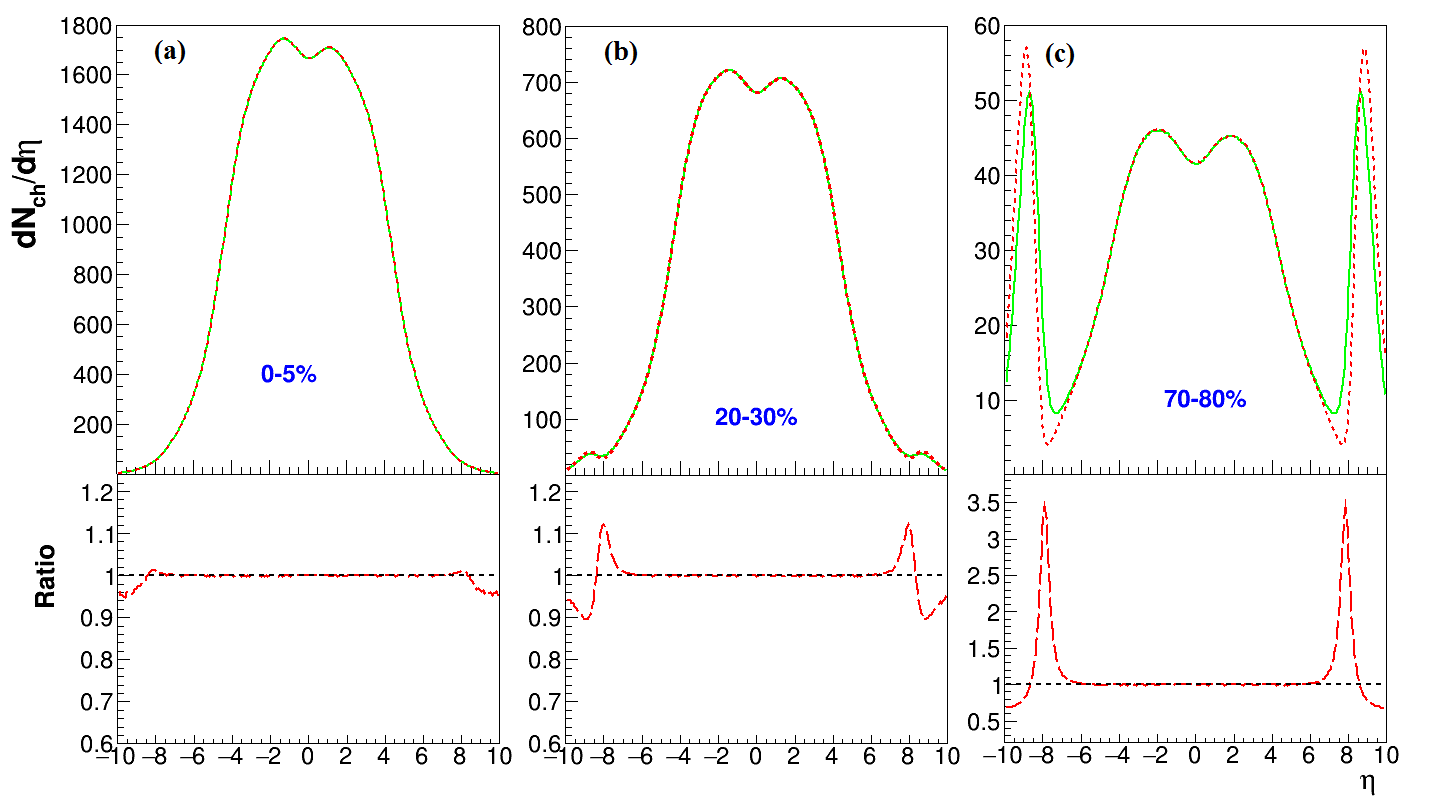}
 		\caption{(COLOR ONLINE)
 			Pseudorapidity density of charged particles in $Pb+Pb$ collisions at $\sqrt{s_{NN}}=2.76$ TeV and different centrality intervals: (a) for central (0-5\%), (b) for semi-peripheral (20-30\%) and, (c) for peripheral (70-80\%) interactions. The short dashed and solid lines denote  ImHIJING/Secondary calculations without and with energy-momentum conservation constraints, respectively. The lower part in each panel depicts the ratios of $\frac{dN_{ch}}{d\eta}$ results with to those without energy-momentum conservation constraints (long dashed lines).
 		}
 		\label{f6}
 	\end{center}
 \end{figure*} 

By inspection of Fig.\ref{f1}, one can conclude that the ImHIJING/Secondary results with the two different set of parameters, for different centrality at mid-pseudorapidity, are independent of the assumed form of parton shadowing, which reflects the delicate balance between hard and soft processes. Notice that the maximum value of parton shadowing for the two sets  is smaller than HIJING 2.0 (A MultiPhase Transport AMPT \cite{25}) estimate of $s_{q(g)}=0.2-0.23$ \cite{2} ($s_{q(g)}=0.16-0.17$ \cite{26})  which indicates the importance of using the most precise MSTW2009 tabulated form of PDFs, that constrain the gluon (and quark) distributions in the domain $0.01\leq x \leq0.5$, and kinematic constrains on primary interacting nucleons. Also, the AMPT model are shown to be roughly consistent with the experimental data for central collisions, while for peripheral collisions ($<N_{part}>$ $<169)$ the model calculation overestimates ALICE data \cite{26, 27}. 
 
One cannot conclude that the assumed form of parton shadowing is correct without studying both the shape of $\frac{dN_{ch}}{d\eta}$ and $p_{T}$-charged particles spectra. Indeed, as one can see in Fig.\ref{f2} the ImHIJING/Secondary results with set 1 (long-dashed line) cannot describe the overall shape of $\frac{dN_{ch}}{d\eta}$. On the other hand, the model results with set 2 (thick lines) agree with the measured LHC data (see Figs.\ref{f2} and \ref{f3}). Thus, in what follows we will study the kinematic constrain effects using only set 2. 
                       
In Fig.\ref{f2}, we study the influence of final state interactions of large $p_{T}$ jets with the dense nuclear medium, created during $Pb+Pb$ collisions at $\sqrt{s_{NN}}=2.76$ TeV along the transverse direction $x$, on the $\frac{dN_{ch}}{d\eta}$ yield.  The results obtained by  ImHIJING/Secondary are compared to the 0-5\% most central events. The final state interactions scenario of HIJING 1.0 code is adopted \cite{1}, where a part of the jet energy $\Delta E$ is transferred as a gluon kink to the other string which the jet interacts with. A cut off parameter $\epsilon=4.5$ GeV/fm is defined below which the jet can not lose energy anymore via interactions with the medium. Note that the value of the cut off is taken the same as the one for jet production $p_{0}$. As one can see (thin line), turning off jet quenching $(dE/dx=0)$ results in a reduction of total charged particles as much as 25\% at mid-pseudorapidity.  By increasing the final state interactions, when the jet energy loss $(dE/dx)$ is less than $\epsilon$, the ImHIJING/Secondary predicts the level at mid-pseudorapidity but fails to reproduce the over all shape of $\frac{dN_{ch}}{d\eta}$. Note that, the nearly flat $\frac{dN_{ch}}{d\eta}$ at $|\eta\rvert\leq2$ has also been observed \cite{9} with models employing final state parton cascade, as implemented in AMPT model \cite{25}. 

This implies that one cannot conclude the validity of initial conditions of HIJING type models without reproducing the whole shape of $\frac{dN_{ch}}{d\eta}$.  At $dE/dx=\epsilon$, final state interactions yield a shoulder at  $\lvert \eta \rvert=2$ that are found to be in good agreement with the measured $\frac{dN_{ch}}{d\eta}$ distribution, indicating that the Pb+Pb collisions at $\sqrt{s_{NN}}=2.76$ TeV are transparent.

In Fig.\ref{f3} we check the effect of changing the energy loss parameter on the invariant transverse momentum $(p_{T})$ spectra at mid-pseudorapidity $(\lvert \eta \rvert<0.8)$ for the reaction under study. As expected, the ImHIJING/Secondary calculation  with  
$ dE/dx=0 $ overpredicts the $p_{T}$-spectra at  $p_{T}>3$ GeV/c for  both  central and peripheral collisions (thin lines).  At  $dE/dx=\epsilon$, we find that the ImHIJING/Secondary spectra are clearly suppressed at $p_{T}>3$ GeV/c due to jet quenching, in accordance with ALICE experimental data \cite{23}.  We also see that the quenching effect is larger for central than peripheral collisions, as it should be.
 
It is worthwhile noting  that the AMPT model (which includes both jet quenching and final state partonic interactions) reasonably describes the charged particles $p_{T}$-spectra for $p_{T}<1$ GeV/c but gives smaller values for larger $p_{T}$ \cite{26, 27}. This indicates that the recent versions of HIJING type models show more significant quenching of final state hard scattered partons than seen by the LHC data. Thus, although the present calculation from the HIJING model is lacking the process of induced gluon radiation and the accompanied energy loss from leading parton in QCD medium, our results with the simple jet quenching mechanism agree with the measured LHC data.

Next, in Fig.\ref{f4}, we compare both the ImHIJING/Primary and ImHIJING/Secondary results with the LHC data of $\frac{dN_{ch}}{d\eta}$ per  centrality class (from 0-5\% to 70-80\%) for the reaction under study. The ImHIJING/Primary and ImHIJING/Secondary calculations differ in the kinematic constrains on secondary interacting nucleons. As one can see,  ImHIJING/Primary yields a similar shape to the measured results, but overestimates the level at mid-pseudorapidity $(\lvert \eta \rvert \le 2)$ with increasing centrality (small dashed lines). On the other hand, calculations with ImHIJING/Secondary are a good fit for the whole centrality dependence of $\frac{dN_{ch}}{d\eta}$ (solid lines). 
It should be pointed out that, the UrQMD, a Color Glass Condensate-type \cite{28}  and AMPT (which uses HIJING2.0 as initial conditions) models, fail to reproduce the overall level and shape of ALICE data \cite{9}. This may imply that the ImHIJING constrains of both primary and secondary interactions are more consistent with LHC data.

It is interesting to note that, the introduction of kinematic constrains on primary and secondary interacting nucleons  enhances  hadron production at large rapidity $\lvert \eta \rvert >8$,  see Fig.\ref{f4}. This effect can be quantified by analyzing the pseudorapidity density of the main charged species such as protons, $\pi^{\pm}$ and $K^{\pm}$, as shown in Fig.\ref{f5}. It is clear from the figure that the enhanced production of charged particles at  $\lvert \eta \rvert >8$ is mainly due to interactions between protons at the spectator parts of the collision (long dashed lines). This implies that kinematic constrains result in a correct geometrical treatment of $Pb+Pb$ collisions at LHC energies, especially at large rapidity $\lvert \eta \rvert >8$ for centrality intervals starting from 20-30\% $(b\geq R_{A})$.
                            
Finally, in order to study the effect of imposing energy-momentum conservation constrains on secondary interacting nucleons,  we compare in Fig.\ref{f6} the ImHIJING/Secondary results of $\frac{dN_{ch}}{d\eta}$ for central (0-5\%), semi-peripheral (20-30\%) and peripheral (70-80\%) centrality classes. The ImHIJING/secondary calculations are performed with (solid lines) and without (short-dashed lines) energy-momentum conservation constraints. Due to inclusion of energy-momentum conservation, the predicted $\frac{dN_{ch}}{d\eta}$ is quite sensitive to kinematic constrains in the projectile/target pseudorapidity region. In particular, we clearly see a rise in the ratios of  $\frac{dN_{ch}}{d\eta}$ results with energy-momentum conservation constraints to those without at  $6<\lvert \eta \rvert <8$ before the sharp fall at $\lvert \eta \rvert >8$ as the centrality decreases, though the ratios at mid-pseudorapidity regions are the same. 

The results presented in this paper could be interpreted as follows. The secondary interactions induces nuclear modification of nucleons that are involved in the primary interactions. More specifically, nucleons taking part in the primary interactions suffer energy loss due to cascading with other non-interacting ones, and the remaining energy is used to produce jets or excited strings according to  ImHIJING, that in turn should influence the description of $\frac{dN_{ch}}{d\eta}$ as a function of collision centrality.

\section{\label{sec:level4}SUMMARY AND CONCLUSIONS}
The improved HIJING (ImHIJING) calculations, that with recent  MSTW2009 parton distribution functions determined from global analysis of hard scattering data, have been performed for  the dependence of charged particle pseudorapidity density $\frac{dN_{ch}}{d\eta}$ on collision centrality (from 0-5\% to 70-80\%) in $Pb+Pb$ collisions at $\sqrt{s_{NN}}=2.76$ TeV. To assess the effect of kinematic constrains on interacting nucleons in heavy ion collisions, we incorporated the collective cascade recipe. We found that the effect changes  the whole $\frac{dN_{ch}}{d\eta}$ yield as a function of collision centrality. In particular, the mid-pseudorapity region $(|\eta|<2)$ of $\frac{dN_{ch}}{d\eta}$ is  $\sim5$\% reduced, independent of collision centrality, by the kinematic constrains on primary interacting nucleons, in agreement with the measured LHC data. Concerning the pseudorapidity region of the projectile/target $(\lvert \eta \rvert>8)$, an enhanced production of hadrons is clearly observed. The kinematic effects at $\lvert \eta \rvert>8$ are shown to be due to the interactions of protons at peripheral collisions. 
  
Therefore, the inclusion of kinematic bias, due to energy-momentum conservation constrains, on interacting nucleons in nucleus-nucleus collisions induces not only nuclear modification of nucleons (nucleon shadowing), that are involved in primary interactions, but also a correct geometrical treatment of the interactions at the projectile/target spectator regions. Such kinematic constrains should be implemented  in microscopic transport approaches for a better account of the initial conditions in LHC energy heavy-ion collisions.\\

\section*{\label{sec:level}ACKNOWLEDGMENTS}
The authors would like to thank Prof. V.V. Uzhinskii for checking the ImHIJING/Secondary code. Kh. A.-W. would like to thank the members of GEANT4 hadronic group for the hospitality and advice during his visits to CERN. The authors would also like to thank the referee for the comments that improved the quality of the paper.

\end{document}